\documentclass{aa}  
\def\mean#1{{\langle}#1{\rangle}}
\newif\ifAMStwofonts 
\usepackage{graphicx}
\usepackage{txfonts}

\begin{document}

\title{Oligarchic planetesimal accretion and giant planet formation}
  
   \author{A. Fortier
          \inst{1,2}\fnmsep\thanks{Fellow of CONICET}
          \and
	  O. G. Benvenuto
	  \inst{1,2}\fnmsep\thanks{Member of the Carrera del 
	  Investigador Cient\'{\i}fico, Comisi\'on de Investigaciones 
	  Cient\'{\i}ficas de la Provincia de Buenos Aires, Argentina.
	  E-mail: obenvenu@fcaglp.unlp.edu.ar}
	  \and
          A. Brunini\inst{1,2}\fnmsep\thanks{Member of the Carrera del Investigador 
	  Cient\'{\i}fico, Consejo Nacional de Investigaciones  Cient\'{\i}ficas
	   y T\'ecnicas, Argentina. 
	  E-mail: abrunini@fcaglp.unlp.edu.ar}
          }

   \offprints{A. Fortier}

   \institute{Facultad de Ciencias Astron\'omicas y Geof\'{\i}sicas, 
   Universidad Nacional de La Plata, Paseo del Bosque s/n (B1900FWA) La Plata, 
   Argentina\\
              \email{afortier@fcaglp.unlp.edu.ar}
         \and
	      Instituto de Astrof\'{\i}sica de La Plata, IALP, CONICET-UNLP, 
	      Argentina\\
           }

   \date{Received ; accepted }

  \abstract
  {}
   {In the context of the core instability model, we present calculations of 
   \emph{in situ\/} giant planet formation. The oligarchic growth regime of
   solid protoplanets is the model adopted for the growth of the core. 
   This growth
   regime for the core has not been considered before in full evolutionary
   calculations of this kind.}
  {The full differential equations of giant planet formation were numerically 
   solved with an adaptation of a Henyey--type code. The planetesimals accretion
   rate was coupled in a self--consistent way to the envelope's evolution.} 
  {We performed several simulations for the formation of a Jupiter--like object 
 by assuming various surface densities for the protoplanetary  disc and two 
 different sizes 
 for the accreted planetesimals. We first focus our study on the atmospheric 
 gas drag that the incoming planetesimals suffer. We find that 
 this effect gives rise to a major enhancement on the effective capture 
 radius of the protoplanet, thus leading to an average timescale reduction of 
 $\sim$ \mbox{30\%} -- \mbox{55\%} 
 and ultimately to an increase by a factor of 2 of 
 the final mass of solids accreted as compared 
 to the situation in which drag effects are neglected. In addition, we also 
 examine the importance of the size of accreted planetesimals on the whole 
 formation process. With regard to this second point,
 we find that for a swarm of planetesimals 
 having a radius of 10 km, the formation time is a factor 2 to 3 shorter
 than that of planetesimals of 100 km, the factor depending on the surface
 density of the nebula. 
 Moreover, planetesimal size does not seem to have a significant 
 impact on the final mass of the core.}
 {}


   \maketitle
%

\section{Introduction}
\label{sec:intro}

Although the giant planets of the Solar System are known from centuries ago and
 the existence of other planetary systems harbouring Jupiter--like 
objects was confirmed in 1995 (Mayor \& Queloz \cite{mayor}), the mechanism of 
giant planet formation is still not solved. In fact, two major models
currently coexist: the core accretion model --- alternatively referred to as 
the nucleated instability model --- and the gas instability model (for a 
detailed review on the theory of giant planet formation see, e.g.,  Wuchterl 
et al. \cite{wuchterl}; Lissauer \& Stevenson \cite{lissauer}; Durisen et al. 
\cite{durisen}).
  
The core accretion model assumes that giant planets form, roughly speaking, 
in a two--step process (e.g. Mizuno \cite{mizuno}; Bodenheimer \& Pollack 
\cite{bodenheimer}). First, analogous to terrestrial planets, a 
solid core is formed by coagulation of planetesimals. Once the core is massive 
enough to capture a significant amount of gas from the surrounding nebula, 
the characteristic extensive envelope of these objects forms. 
Gravitational gas binding occurs mainly in two 
different regimes (Pollack et al. \cite{pollack}). Initially, gas 
accretion proceeds relatively quietly. This stage ends when the mass of the 
envelope is comparable to the mass of the core. When this critical mass is 
reached, a gaseous runaway growth completes the formation of the planet, 
accreting large amounts of gas in a very short time. 

Evidence that the giant planets of the Solar System have a core approximately
a dozen times the mass of Earth, favours the nucleated instability scenario 
over the gas instability hypothesis. However, one of the problems that still 
remains open with the core 
accretion model is how to reconcile the formation timescales evolving from 
numerical simulations with those obtained by observations of circumstellar discs
(e.g., Haisch et al. \cite{haisch}; Chen \& Kamp \cite{chen}). Observational 
evidence restricts the lifetime of protoplanetary discs to less than $10^7$ 
yr. This fact puts a limit on the whole process: giant planets should be 
formed before the dissipation of the disc. After the pioneering work of Pollack 
et al. (\cite{pollack}), in which detailed calculations of giant planet 
formation were performed, 
efforts have been made to solve this problem. In particular, different 
scenarios that could alleviate the timescale issue have been explored, 
including: planet migration (Alibert et al. \cite{alibert}),
grain opacity reduction (Podolak \cite{podolak}; Hubickyj et al. 
\cite{hubickyj}), and core formation in the centre of an anticyclonic vortex 
(Klahr \& Bodenheimer \cite{klahr}) among others. To this end, most authors 
who adopt a time--dependent solid accretion rate for the core (Pollack et al. 
\cite{pollack}; Hubickyj et al.  \cite{hubickyj}; Alibert et al. \cite{alibert})
prescribe a rapid one (Greenzweig \& Lissauer \cite{greenzweig})
which leads to the formation of a massive core in a few hundred thousand 
years. This accretion rate slows down when most of  the material within the 
feeding zone is swallowed by the embryo. With this core accretion rate, 
the formation timescale of a core of about 10 $\mathrm{M_{\oplus}}$ usually 
only represents a
 small fraction of the whole formation process (Pollack et al. \cite{pollack}). 
However, N--body simulations show that a solid embryo with the Moon mass, or 
even smaller, could gravitationally perturb the swarm of planetesimals 
around it, heating the disc and decreasing solid accretion long before the 
isolation mass is achieved (Ida \& Makino \cite{ida}). The runaway 
growth of the large planetesimals (the first stage in protoplanet building) 
then switches to a slower, self--limiting mode. 
 This regime is known as  ``oligarchic growth'' (Kokubo \& Ida 
\cite{kokubo1}, \cite{kokubo2}, \cite{kokubo3}; Thommes et al. \cite{thommes}). 
 Numerical and semi--analytical calculations were made to estimate 
the formation timescale of a solid protoplanet via the oligarchic growth 
(Ida \& Makino \cite{ida}; Thommes et al. \cite{thommes}). However, no full 
evolutionary calculations of gaseous giant planet formation prescribing this 
accretion rate have been made to date.

In the frame of the core instability model, this paper is intended to study
the \emph{in situ\/} formation of a protoplanet in a circular 
orbit around the Sun, at the current position of Jupiter.
To this end, the numerical code developed by Benvenuto \& Brunini 
(\cite{bb}) for self--consistent calculation of giant planet formation and 
evolution has been updated.
Motivated by the work of Thommes et al. (\cite{thommes}), where the oligarchic 
regime was applied to growing solid embryos in the absence of an atmosphere, we
generalise this accretion model in order to adopt it as the time--dependent 
accretion rate for the core. 
The accretion rate employed in Thommes et al. (\cite{thommes}) is modified 
to take into account 
the gas drag that incoming planetesimals suffer due to the presence of
the atmosphere, as it enlarges the protoplanet's capture cross--section.
Inaba \& Ikoma (\cite{inaba}) derived a semi--analytical core 
accretion rate for
 a protoplanet surrounded by a gaseous atmosphere and estimated the enhancement 
in the collisional cross--section it produced. In their study, the structure of 
the atmosphere is calculated only for certain core masses and not as a result of
 an evolutionary sequence of models.
In our study, the solids accretion rate is coupled to the gas accretion rate, 
leading to a self--consistent evolutionary calculation of the envelope's 
structure. Hence, a more accurate estimation of the enlargement of the effective 
accretion cross--section of the protoplanet can be made. 

This paper is organised as follows: In Sect. \ref{sec:procedure}  we describe 
the improvements in our numerical code, placing particular emphasis on the 
treatment of the mass accretion rate of planetesimals onto the core. 
In connection with the growth of the core,
 we summarise some conceptual aspects and the main results 
regarding the oligarchic growth of a solid protoplanet.
In Sect. \ref{sec:results} we present a simulation which is compared to
results obtained by other authors which use a different planetesimals accretion 
rate. This is intended to show the impact that the selected core accretion model
 has on the formation of a giant planet. We then present the results  
arising from running our code
for simulations of the formation of a Jupiter--like planet for several 
protoplanetary disc surface densities. 
We also explore the relevance of planetesimal size,
comparing runs for a swarm of planetesimals having a radius of 10 and 100 km. 
Section \ref{sec:conclusion} is devoted to discussing our results in connection 
with other related studies and to summarising the main conclusions of our study.

\section{Outline of the overall model}
\label{sec:procedure}

The description of the numerical model developed for the calculation of giant 
planet formation and evolution was introduced in Benvenuto \& Brunini 
(\cite{bb}). Since then, several changes have been made to the 
code. In this section we will summarise the main improvements.

\subsection{The protoplanetary disc}

\label{sec:disc}
We characterise the protoplanetary disc with only three parameters: the 
temperature profile, the solids surface density and the gas volume density, 
under the hypothesis of the model proposed by Hayashi (\cite{hayashi}).

We consider a power law for the temperature profile
\begin{equation}
\label{eq:temp}
T \, (a)=\alpha T_\mathrm{eff} \, (1\; \mathrm{AU}) \, T_\mathrm{eff}^{\star} 
\, a^{-q},
\end{equation}
where $a$ is the distance to the central star, $T_\mathrm{eff}^{\star}$ is the 
effective temperature of the central star, relative to that of the Sun 
(it is a dimensionless magnitude and in the cases considered in this article
 $T_\mathrm{eff}^{\star}= 1$), 
$T_\mathrm{eff} \mathrm{(1 \; AU)}$ is the effective temperature of the disc 
at 1 AU which was fixed at 280 K 
(the current temperature in the Solar System at that location), $q$ is 
the temperature index (here $q = 1/2$) and $\alpha$ is a factor that compensates
  for the fact that the central star was more luminous in the early epoch 
($\alpha$ was elected to be 1.08). The snow line, $\,a_\mathrm{snow}$, is 
located at \mbox{$T=170$ K}, which corresponds in this model to 
$a= 3.16 \; \mathrm{AU}$.

The definition of the Minimum Mass Solar Nebula (MMSN) assumes a solids surface 
mass density profile that is a power law
\begin{equation}
\label{eq:sigma1}
\Sigma \, (a)= \eta_\mathrm{ice}\;\sigma_\mathrm{z} \, (1\;
\mathrm{AU})\;a^{-p},
\end{equation}
where we fixed
$\sigma_\mathrm{z} \, \mathrm{(1 \;AU)} = 10 \; \mathrm{g \;cm^{-2}}$, $p=3/2$ 
and $\eta_\mathrm{ice}$ is the enhancement factor beyond the snow line,
\begin{displaymath}
\label{eq:eta}
\eta_\mathrm{ice}=\left\lbrace
\begin{array}{cc}
1 & \mathrm{if} \; a<a_\mathrm{snow}\\
4 & \, \mathrm{if} \; a>a_\mathrm{snow} .
\end{array}
\right.
\end{displaymath}

Although in this equation the assumption of a uniformly spread 
mass of solids is implicit, for the calculation of giant planet formation it is 
also necessary to adopt a planetesimal mass and radius (see the following 
sections). In this study we consider that the planetesimal bulk density, 
$\rho_m$, is constant and equal to $1.5 \; \mathrm{g \; cm^{-3}}$ and that 
all planetesimals in the disc are of the same size.

The gas volume density of the solar nebula is given by
\begin{equation}
\label{eq:rho}
\rho \, (a)= \sigma_\mathrm{g}(1\; \mathrm{AU})\;\frac{a^{-b}}{2H}, 
\end{equation}
where $\sigma_\mathrm{g}$ (1 AU) is the gas surface density at 1 AU taken 
to be $2 \times 10^3 \; \mathrm{g \; cm^{-2}}$, $b=3/2$ and $H$ is the gas 
disc scale height, 
\begin{equation}
\label{eq:altura}
H \, (a)=0.05 \; a^{5/4} \times (1.5\times 10^{13} \mathrm{cm}).
\end{equation}
Note that $a$ is always in AU but $H$ must be in $\mathrm{cm}$ for 
$\rho \,(a)$ to be in $\mathrm{g \; cm^{-3}}$ giving the 
factor $ 1.5\times 10^{13} \; \mathrm{cm}$.

The main features of the MMSN at 5.2 AU are reported in \mbox{Table 
\ref{table:nebula}}.

\begin{table}
\caption{ Characteristics of the Minimum Mass Solar Nebula at 5.2 AU.}
\label{table:nebula}
\centering
\begin{tabular}{p{4cm} l}
\hline
\hline
\\                      
Temperature   & 133 $\mathrm{K}$ \\
Solids surface density    & 3.37 $\mathrm{g\;cm^{-2}}$ \\
Gas volume density   & $1.5 \times 10^{-11} \mathrm{g\;cm^{-3}}$ \\

\hline 
\end{tabular}
\end{table}

\subsection{The core}
\label{sec:core}

\subsubsection{The oligarchic growth of protoplanets}
\label{sec:oligarchic}

Assuming that kilometre sized planetesimals can be formed in a protoplanetary 
disc, it is generally accepted that the first seeds for rocky
protoplanets emerge through planetesimal accretion. The early stage in
protoplanetary growth is the so--called ``runaway growth'' (Greenberg et al.
\cite{greenberg}; Kokubo \& Ida \cite{kokubo0}). Runaway growth can be 
summarised as follows:
random velocities of large planetesimals are smaller than those of small
planetesimals due to dynamical friction. This fact favours the gravitational
focusing of large planetesimals which leads them to grow in a runaway
fashion. These objects rapidly become more massive than the rest of the
planetesimals in the swarm and the first planetary embryos appear in the disc.
During the runaway stage, both the growth rate of a protoplanet and the mass
ratio of a protoplanet and planetesimals increase with time. 

Ida \& Makino (\cite{ida}) investigated the post--runaway evolution of 
planetesimals' eccentricities and inclinations due to scattering by a 
protoplanet. When runaway embryos become massive enough to affect
planetesimals' random velocities, the growth regime switches to a slower one. 
Ida \& Makino (\cite{ida}) called this stage ``the protoplanet--dominated 
stage'',  
as protoplanets are now responsible for the larger random velocities of
planetesimals which, in turn, decreases the growth rate of protoplanets.
In their study of this regime, the authors performed 3D N--body simulations to
investigate the evolution of eccentricities and inclinations of a system of 
equal--mass planetesimals and one protoplanet. Protoplanet--planetesimal and
planetesimal--planetesimal interactions were both taken into account, but no
accretion was considered. Ida \& 
Makino (\cite{ida}) found that planetesimals are strongly scattered by the
protoplanet and part of the energy they acquire during the perturbation is 
distributed afterwards among other planetesimals. The excited planetesimals 
determine a ``heated region'' around the protoplanet twice the width of the
feeding zone of the protoplanet. In this region, the eccentricities $e_m$ and
inclinations $i_m$ of planetesimals are highly perturbed and, as 
random velocities of planetesimals are well approximated by 
\begin{equation}
\label{eq:vrel}
v_\mathrm{rel} \simeq \sqrt{\mean{e_m^2}^{1/2}+\mean{i_m^2}^{1/2}}
\,v_\mathrm{k},
\end{equation} 
 the increase in $e_m$ and $i_m$ directly implies an increase in random
velocities ($v_\mathrm{k}$ is the Keplerian velocity, $\mean{e_m^2}^{1/2}$ 
($\mean{i_m^2}^{1/2}$) is the planetesimals RMS eccentricity (inclination)).

Ida \& Makino (\cite{ida}) then proposed a two--step growth for the protoplanets.
In the first stage, protoplanets grow in a runaway fashion in the usual sense: 
the stirring among planetesimals is dominated by planetesimals themselves and
relative velocities are low, favouring the rapid growth of the embryos. When
a protoplanet is massive enough to perturb the surrounding planetesimals, the
system enters the protoplanet--dominated stage. The numerical simulations show
that this stage occurs in the dispersion--dominated regime and the relationship
between mean planetesimals' eccentricities and inclinations is 
$\mean{e_m^2}^{1/2} / \mean{i_m^2}^{1/2} \simeq 2$.
 In this second stage, the growth
rate of the protoplanet decreases as a consequence of the greater relative
velocities between the protoplanet and planetesimals. However, the mass
ratio of the protoplanet and planetesimals still increases with time. 

Ida \& Makino (\cite{ida}) also derived, using a semi--analytical model, the 
condition for the dominance of the protoplanet--planetesimal scattering over the
planetesimal--planetesimal scattering:
\begin{equation}
\label{eq:condicion}
2\;\Sigma_\mathrm{M}\; M > \Sigma\; m,
\end{equation}
where $M$ is the mass of the solid embryo, $m$ is the effective 
planetesimal mass, 
$\Sigma$ is the surface mass density of the planetesimal disc and $\Sigma_
\mathrm{M}$ is defined as
\begin{equation}
\label{eq:sigmaM}
\Sigma_\mathrm{M} = \frac{M}{2 \pi a \Delta a}, 
\end{equation}
with $2 \pi a \Delta a$ the area of the ``heated region''. For a standard solar
nebula, this condition can be translated into a relationship between $M$ and
$m$ which depends on the density profile of the disc, the semi--major axis $a$ 
and the planetesimal mass. For the low--mass nebula model, the transition occurs
when $M/m \simeq 50-100$. This condition was also corroborated by the authors 
through N--body simulations. Therefore, the second stage of the protoplanet
growth begins when the mass of the protoplanet is still much smaller than the
total mass of planetesimals in the feeding zone.

Ida \& Makino (\cite{ida}) estimated the characteristic growth time (the
mass--doubling time) for the protoplanet (assuming a constant solids surface 
density) : 
\begin{equation}
\label{eq:tgrow}
T_{grow} \equiv \left( \frac{1}{M} \frac{dM}{dt} \right)^{-1} \propto M^{-1/3}
\mean{e_m^2}.
\end{equation}
To estimate $\mean{e_m^2}$, they assumed an equilibrium state where the
enhancement due to viscous stirring is compensated by the damping due to the
nebular gas drag. As a result, in the first stage where the dominant perturbers 
are the planetesimals, $\mean{e_m^2} \propto 
m^{8/15}$ (note that it does not depend on the mass of the protoplanet), while
in the second stage $\mean{e_m^2} \propto m^{1/9}M^{2/3}$. Hence, in the first
stage $T_{grow} \propto M^{-1/3}$ so protoplanets grow very fast and the ratio
$M/m \simeq 50-100$ is reached in a negligible time. In the second stage, the 
dominant
perturber is the protoplanet and $T_{grow} \propto M^{1/3}$, thus the growth of
the protoplanet gradually becomes slower as its mass increases. This type of
time dependence is characteristic of the orderly growth (the presumed final
growth regime for the terrestrial planets).
However, the growth regime in the protoplanet--dominated stage 
is much shorter than the orderly growth because protoplanets
grow by accretion of planetesimals and not through collisions of comparable sized
bodies where dynamical friction is absent. The three growth regimes (the runaway
regime, the 2--step regime and the orderly regime) are schematically illustrated
in Fig. 11 of Ida \& Makino (\cite{ida}).

Kokubo \& Ida (\cite{kokubo1}) performed 3D N--body calculations to study the
post--runaway accretion scenario. Their simulations start with two protoplanets 
that grow by planetesimal accretion. In the first stage of accretion,
 protoplanets grow in a runaway fashion. Protoplanetary runaway growth breaks
down when $M \sim 50 \, m$, as predicted by the semi--analytical theory of 
Ida \& Makino (\cite{ida}). Protoplanets subsequently grow keeping a
typical orbital separation of $10 \, R_\mathrm{H}$, where $R_\mathrm{H}$ is the
 Hill radius of a protoplanet, 
\begin{equation}
\label{eq:hill}
R_\mathrm{H}= \left(\frac{M}{3M_{\star}}\right)^{1/3}a
\end{equation}
with $a$ the distance to the central star.
The orbital repulsion is a consequence
of the coupling effect of scattering between large bodies and dynamical friction
from the swarm of planetesimals. Protoplanets continue their growth keeping 
their mass ratio close to unity, but 
the mass ratio between protoplanets and planetesimals continues to increase
with time. Thus, protoplanets grow oligarchically, as no substantial 
accretion between planetesimals themselves is found. The mass distribution then 
becomes bimodal: a small number of protoplanetary embryos and a large number of
small planetesimals shape the protoplanetary disc. 
 Kokubo \& Ida (\cite{kokubo1}) coined this growth regime
``oligarchic growth'', `in the sense that not only one but several protoplanets
dominate the planetesimal system'.

Kokubo \& Ida (\cite{kokubo2}) investigated, through 3D N--body simulations, the
growth from planetesimals to protoplanets including the effect of the nebular
gas drag. They confirmed the existence of an initial runaway phase in
protoplanetary growth and a second stage of oligarchic growth of protoplanets, in
the same sense as in Kokubo \& Ida (\cite{kokubo1}). Also, when analysing
the evolution of the RMS eccentricity and inclination of the planetesimal system
during the oligarchic growth stage 
they found that their results agree with those predicted by the semi--analytical
theory of Ida \& Makino (\cite{ida}) for their second stage.

\subsubsection{The growth of the core }
\label{sec:acc_rate}
In the present version of the code we have introduced a time--dependent
planetesimal accretion rate. In this kind of calculation most authors 
(Pollack et al. \cite{pollack}; Hubickyj et al. \cite{hubickyj}; Alibert et al.
\cite{alibert}) usually prescribe that obtained by Greenzweig \& Lissauer
(\cite{greenzweig}) which assumes a rapid growth regime for 
the core. Instead, we adopt that corresponding to the oligarchic growth of 
Ida \& Makino (\cite{ida}); a slower accretion rate that still has not been 
 explored with a self--consistent code for giant planet formation.

The condition for the dominance of the oligarchic growth over the (previous) 
runaway growth of a protoplanet, $M/m \simeq 50-100$, was derived
semi--analytically by Ida \& Makino (\cite{ida}). In all the cases of interest 
for this study, the cross--over mass is
very low --- i.e. some orders of magnitude below the Earth mass (Thommes et 
al. \cite{thommes}).
Due to the initial runaway regime, the cross--over mass is reached in a 
negligible time and thus, the formation time of a giant planet's core is 
almost entirely regulated by the oligarchic growth.
For this reason, we prescribe the oligarchic growth for the core
since the very beginning of our simulations.

In the dispersion--dominated regime, a solid embryo growth rate is well 
described by the particle--in--a--box approximation (Safranov, \cite{safranov}), 

\begin{equation}
\label{eq:accrete}
\frac{dM_\mathrm{c}}{dt} \simeq F\frac{\Sigma}{2h}\pi \, R_\mathrm{eff}^2 \, 
v_\mathrm{rel}, 
\end{equation}
where $M_\mathrm{c}$ is the mass of the solid protoplanet (in our case, the core
of the giant planet), $h$ is the 
planetesimals' disc scale height, $R_\mathrm{eff}$ is the effective capture
radius and $F$ is a factor 
introduced to compensate for the underestimation of the accretion rate by a 
two--body algorithm when considering the velocity dispersion of a population of 
planetesimals  modelled by a single eccentricity and inclination equal to the 
RMS values. $F$ is estimated to be $\sim 3$ (Greenzweig \& Lissauer 
\cite{greenzweig}).

Due to gravitational focusing, the effective capture radius of a 
protoplanet, $R_\mathrm{eff}$, is larger than its geometrical radius 
\begin{equation}
\label{eq:reff}
R_\mathrm{eff}^2= R_\mathrm{c}^2 \left( 1+\left(\frac{v_\mathrm{esc}}
{v_\mathrm{rel}} \right)^2 \right), 
\end{equation} 
with $R_\mathrm{c}$ the geometrical radius of the solid embryo, 
$v_\mathrm{esc}$ the escape velocity from 
its surface and $v_\mathrm{rel}$ the relative velocity between the protoplanet 
and planetesimals,  
\begin{equation}
\label{eq:vrel}
v_\mathrm{rel} \simeq \sqrt{e^2+i^2}\,a\,\Omega_\mathrm{k} , 
\end{equation} 
where, hereafter,  $e=\mean{e_m^2}^{1/2}$ ($i= \mean{i_m^2}^{1/2}$ ) is 
planetesimals RMS 
eccentricity (inclination) with respect to the disc ($e$, $i$ $<<$ 1) and 
$\Omega_\mathrm{k}$ is the Keplerian angular velocity. We apply the 
approximations $i \simeq e/2$ and $h \simeq ai$. 
Following Thommes et al. (\cite{thommes}), we adopt for $e$ the equilibrium 
expression that is deduced for the case when gravitational perturbations due 
to the protoplanet are balanced by dissipation due to gas drag,
\begin{equation}
\label{eq:excent}
e=\frac{1.7 m^{1/15}M^{1/3}\rho_m^{2/15}}{\beta^{1/5}C_\mathrm{D}^{1/5}
\rho^{1/5}M_{\star}^{1/3}a^{1/5}}, 
\end{equation}
where $M$ is the protoplanet mass (here we assume $M$ to be the total mass of 
the proto--giant planet, meaning the core mass plus the envelope mass), 
$\rho_m$ is the planetesimal bulk density, $\rho$ is the gas volume density 
of the protoplanetary disc, $C_\mathrm{D}$ is the drag coefficient 
(dimensionless and of the order of 1), $M_{\star}$ is the mass of the central 
star and 
$\beta$ is the width of the ``heated region'' in units of the Hill radius 
(considering that potentially other embryos are growing as well, $\beta$ is of 
 the order of 10). For calculations that do not assume equilibrium values for 
 the eccentricity and the inclination,
 we refer the reader to Chambers (\cite{chambers}). His results show that 
using equilibrium values for 
$e$ and $i$ is an acceptable approximation when considering embryos of 
$m \ga 10^{-2} \; \mathrm{M_{\oplus}}$, consistent with the initial core mass 
of all our simulations (see Sect. \ref{sec:results}).

However, when the solid embryo is massive enough to
gravitationally bind gas from the surrounding nebula, the presence of  
 this envelope should be considered when calculating the effective capture 
 radius, $R_\mathrm{eff}$. The gaseous envelope modifies the 
trajectory of incoming planetesimals, as they are affected by the gas drag 
 that enlarges the capture radius of the protoplanet. 
In addition to the gravitational focusing, a ``viscous focusing'' due to 
gas drag should be considered.
 
The effective radius $R_\mathrm{eff}$, in the form stated in Eq. (\ref{eq:reff}) 
dominates when the mass of bound gas is negligible. But when the embryo acquires 
enough gas to form a thin atmosphere, its effective radius 
becomes larger and it separates from that calculated previously.

To calculate the effective radius of the protoplanet in the presence of gas 
around the solid embryo we take into account the action of gravity and gas drag
 on the
incoming planetesimals. Consider one planetesimal entering the protoplanet
atmosphere with a velocity $v_\mathrm{rel}$ (Eq. (\ref{eq:vrel})). Its equation
of motion results from the action of gravity 
\begin{equation}
\label{eq:grav}
\vec{f}_\mathrm{G}= - \frac{G M_r m}{r^2} \hat{r}, 
\end{equation}
together with the action of gas drag.
In Eq. (\ref{eq:grav}), $r$ is the radial coordinate from the centre of the 
protoplanet and $M_r$ is the mass contained within $r$.
 For the gas drag force acting on a spherical
 body of radius $r_m$ travelling with velocity $v$, we adopt the Stokes law
\begin{equation}
\label{eq:stokes}
\vec{f}_\mathrm{D}=- \frac{1}{2} C_\mathrm{D} \, \pi \, r_m^2 \, 
\rho_\mathrm{g} v^2 \, \hat{v},
\end{equation}
with $\rho_\mathrm{g}$ the envelope density and $C_\mathrm{D}=1$ (Adachi et al.
\cite{adachi}).
Starting at the external  radius of the protoplanet (see Sect. 
\ref{sec:envelope}, Eq. (\ref{eq:radius}) for its definition), numerical 
integrations of the
equation of motion are performed with an  adaptive Runge Kutta 4 routine (Press
et al. \cite{press}). The initial impact parameter considered for the
integration of the trajectory is the effective radius in the absence of gas (Eq.
(\ref{eq:reff}), hereafter $R^*_\mathrm{eff}$). Note that this is the lowest
possible effective radius of the protoplanet. We calculated the trajectory of the
planetesimal until it  completes one close orbit inside the Hill sphere of the
protoplanet (so it is  always gravitationally dominated by the protoplanet). 
At this point, the planetesimal is considered captured and will definitely 
be accreted after a certain time.  This procedure is repeated for increasing
impact parameters (the relative difference between two consecutive trial impact 
parameters being $5 \times 10^{-4}$) until the resulting orbit is no longer
inside the Hill sphere. In other  words, planetesimals are considered to be
captured when the energy lost by gas drag allows them to complete one close
orbit around the core and this is  completely inside the protoplanet. The
largest impact parameter for which this condition is fulfilled is adopted as the
effective radius for that model. The solids accretion rate is still Eq.
(\ref{eq:accrete}), using this new definition for $R_\mathrm{eff}$. The mass of
accreted planetesimals is added to the mass of the core. It is worth mentioning
that according to Inaba \& Ikoma (\cite{inaba}), a detailed  calculation of
planetesimal ablation does not have a significant influence on the estimation of
$R_\mathrm{eff}$. 

Note that our criterion for the capture of planetesimals and the determination of
$R_\mathrm{eff}$ is different to that employed by Pollack et al.
(\cite{pollack}) where $R_\mathrm{eff}$ is taken as the periapsis altitude of
the outermost bound orbit.

\begin{figure} \centering
\includegraphics[angle=-90, width=0.5\textwidth]{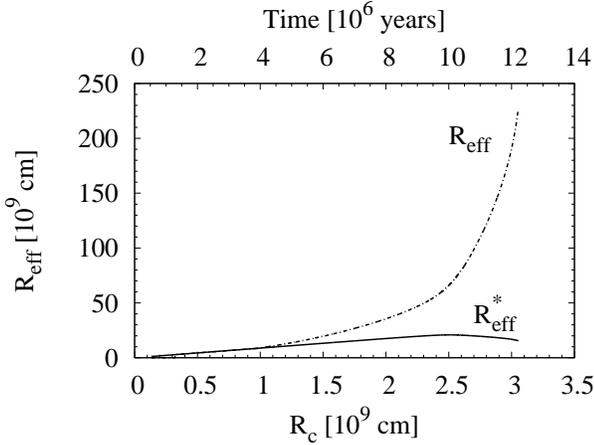}
\caption{Comparison between the protoplanet effective radius for planetesimal
capture. In solid line, $R^*_\mathrm{eff}$ is calculated using Eq. 
(\ref{eq:reff}) and in dashed--dotted line $R_\mathrm{eff}$ is calculated
including the gas drag of the atmosphere. These results were obtained with a
full evolutionary numerical simulation. The protoplanetary disc surface density
corresponds to a  6 MMSN.  The protoplanet is located at 5.2 AU and the radius
of incoming planetesimals is 100 km.}
\label{radcomp}
\end{figure}

We performed a full evolutionary calculation of the formation of a giant planet
in circular orbit around the Sun, with $a=5.2$ AU, and 
immersed in a disc for which solids and gas densities are 
increased by a factor of 6 with respect to the MMSN (see Sect.
\ref{sec:our_results}). 
The effect of gas drag on the protoplanet effective radius is very important, 
as seen in Fig. \ref{radcomp}. 
The effective radius calculated including the gas 
drag separates from the effective radius 
calculated according to Eq.  (\ref{eq:reff}), $R^*_\mathrm{eff}$,
when the elapsed time is $t \sim 4 \; \mathrm{My}$ and the radius of the core is
$R_c \sim \; 10^9 \; \mathrm{cm}$. 
This corresponds to an envelope mass of $\sim 5 \times 10^{-4} 
 \; \mathrm{M_{\oplus}}$ and a core mass of $\sim 2 \; \mathrm{M_{\oplus}}$.
Note that by the end of the formation 
process, $R_\mathrm{eff}$ is more than 10 times larger than 
$R^*_\mathrm{eff}$. Moreover, it enters quadratically in the core 
accretion rate (Eq. (\ref{eq:accrete})), so it will impact on the protoplanet 
formation timescale and on the final core mass (see Sect. \ref{sec:results}). 

The core growth rate (Eq. (\ref{eq:accrete})) depends also on the 
surface mass density of planetesimals, $\Sigma$, a quantity that is neither 
constant in time nor in space. $\Sigma$ is a function of the distance to the 
central star (see Sect. ~\ref{sec:disc}) and also depends on the disc 
evolution. In 
this paper we consider only one effect for the time dependence 
of $\Sigma$: the accretion by the protoplanet. There are other important 
effects to be taken into account for a more realistic calculation of the
planetesimal accretion rate, for example, planetesimal migration 
in the protoplanetary disc due to the nebular gas drag  (Thommes et al. 
\cite{thommes}) and planetary perturbations, for example, planetesimal
ejection away from the feeding zone of the protoplanet (Alibert et al.
\cite{alibert}). However, this paper 
is mainly focused on the coupling of an oligarchic growth regime 
for the core with a full evolutionary calculation of the formation of a
 giant planet, 
so in this first exploration we shall neglect other effects that may contribute
to the time evolution of $\Sigma$.

When considering the time variation of $\Sigma$  
we assume that the growing protoplanet is able to capture planetesimals only 
from its feeding zone, defined as an annulus of $4$ Hill radii (Eq. 
(\ref{eq:hill})) at either side of its orbit.
 The surface density inside the feeding zone at a certain time $t$ is considered
 to be uniform, changing only due to the accretion by the protoplanet 
\begin{equation}
\label{eq:uniform}
\Sigma \equiv \mean{\Sigma}\;(t)= \mean{\Sigma}_0 - \frac{M_\mathrm{c} (t)}
{\pi(a_\mathrm{out}^2(t) - a_\mathrm{int}^2(t))}, 
\end{equation}
where $a_\mathrm{int}(t)$, $a_\mathrm{out}(t)$ are the inner and outer 
boundaries of the feeding zone and $\mean{\Sigma}_0$ is the average of the 
initial surface density  in that region. Adopting this expression for $\Sigma$
tacitly implies that the accretion process is slow enough to guarantee a
uniform distribution of planetesimals inside the feeding zone.

\subsection{Interaction between incoming planetesimals and the protoplanetary 
envelope}
\label{sec:interaction}

On their trajectory to the core, accreted planetesimals interact with the
gaseous  envelope of  the protoplanet exchanging energy.  In this
study, we implemented a simple model to take into
account this  effect (for more detailed models see Podolak et al. 
\cite{podolak2}; Pollack et al. \cite{pollack}; Alibert et al. \cite{alibert}).
 To simplify the situation,  we assume
planetesimals approaching the protoplanet as coming from infinity and entering 
the envelope with velocity $v_\mathrm{rel}$ (Eq. (\ref{eq:vrel})),  
describing afterwards a straight trajectory to the core. This may seem to 
contradict the fact that we calculate the orbits of planetesimals to define 
$R_\mathrm{eff}$, but as mentioned earlier we stop orbit calculation when 
planetesimals complete the first revolution inside the Hill sphere. 
In the future we 
will incorporate a complete calculation of the trajectories to the core, 
together with a self--consistent deposition of planetesimal energy in the 
gaseous layers of the envelope. 

In our simple model of a straight trajectory to the core, the energy of one 
incoming planetesimal of mass $m$ at the top of the atmosphere is
\begin{equation}
\label{eq:ecin}
E=\frac{1}{2} m \, v_\mathrm{rel}^2.
\end{equation}
Gas drag and gravity acceleration are considered here as the two main 
responsible forces for changing the energy of planetesimals throughout the 
envelope. 
The drag force acting on a spherical planetesimal of radius $r_m$ and moving
with velocity $v$ is given by Eq. (\ref{eq:stokes}). The gravitational force 
is calculated as usual with \mbox{Eq. (\ref{eq:grav})}.

Both forces act simultaneously and modify the planetesimal velocity.
The total force acting on a planetesimal is
\begin{equation}
\label{eq:fuerzas}
\vec{f}=\vec{f}_\mathrm{D}+\vec{f}_\mathrm{G}=-m \, \frac{dv}{dt} \, \hat{r}
\end{equation}
and
\begin{equation}
\label{eq:velo}
\frac{dv}{dt}=v \, \frac{dv}{dr}, 
\end{equation}
then
\begin{equation}
\label{eq:speed}
\frac{dv}{dr}= - \frac{C_\mathrm{D} \, \pi \, r_m^2  \, \rho_\mathrm{g} \, v}
{2m} + \frac{G M_r}{r^2  \, v}.
\end{equation}

The kinetic energy lost by a planetesimal due to gas drag $E_{\mathrm{k,}m}$
is transformed into heat, gained by the envelope's shells, 
\begin{equation}
\label{eq:energy}
\frac{dE}{dr}=- \frac{dE_{\mathrm{k,}m}}{dr}= \frac{1}{2} C_\mathrm{D} \, 
\pi r_m^2 \, \rho_\mathrm{g} \, v^2.
\end{equation}
These equations correspond to the energy exchange of one planetesimal with the 
surrounding gas. When considering all incoming planetesimals per unit of time, 
$\dot{M}_\mathrm{c}$, and transforming the differential equations into 
difference equations, the total variation of energy $\Delta E_i$ of the 
envelope's shell $i$  in the time interval $\Delta t$ is
\begin{equation}
\label{eq:deltae}
\Delta E_i= \frac{1}{2} C_\mathrm{D} \, \pi \, r_m^2 \, \rho_g \frac 
{\dot{M}_c}{m} \, v_i^2 \, \Delta r_i  \, \Delta t.
\end{equation}
where $v_i$ (the velocity of planetesimals when entering an envelope's shell 
$i$) is obtained from a discretisation of Eq. (\ref{eq:speed}), 

\begin{equation}
\label{eq:vi}
v_i=v_{i-1} + \left(- \frac{C_\mathrm{D} \, \pi \, r_m^2  \, \rho_\mathrm{g} 
\, v_{i-1}}{2m} + \frac{G M_{r_{i-1/2}}}{r^2_{i-1/2}  \, v_{i-1}}\right) 
\Delta r_{i-1}.
\end{equation}

Finally, when planetesimals reach the core, all their remaining kinetic energy 
is deposited in the adjacent gas shell.

\subsection{The gaseous envelope}
\label{sec:envelope}
The general procedure employed in this study to numerically solve the formation
of a giant planet was 
introduced in a previous paper (Benvenuto \& Brunini \cite{bb}). In the above 
sections we mentioned some of the main improvements introduced to the model 
presented in that work. For the sake of completeness, we summarise here the 
fundamental ideas concerning the solution of the envelope's structure.
 The full
differential equations of giant planet formation are solved with an adaptation 
of a Henyey--type code. Radiative and convective transport are considered, 
employing the Schwarzshild criterion for the onset of convection. The adiabatic 
gradient is adopted for the temperature gradient in the latter case.

The boundary conditions remain the same. As inner boundary conditions, we 
consider the core density to be constant, $\rho_\mathrm{c}=3$ 
$\mathrm{g \; cm^{-3}}$. The mass of the core at time $t$ is calculated as
\begin{equation}
\label{eq:condi1}
M_\mathrm{c}=\frac{4}{3} \, \pi \, \rho_\mathrm{c} \, 
R^3_\mathrm{c}(t).
\end{equation}

The luminosity on the surface of the core is
\begin{equation}
\label{eq:condi2}
L_r(M_\mathrm{c})=0,
\end{equation}
and the velocity is
\begin{equation}
\label{eq:condi3}
v\,(M_\mathrm{c})=\frac{\dot{M}_\mathrm{c}}{4 \, \pi \, R^2_\mathrm{c} 
\, \rho_\mathrm{c}}.
\end{equation}

For the outer boundary conditions, the planetary radius is defined as
\begin{equation}
\label{eq:radius}
R= \mathrm{min} [R_\mathrm{acc}, R_\mathrm{H}], 
\end{equation}
where the accretion radius, $R_\mathrm{acc}$, is given by
\begin{equation}
\label{eq:racc}
R_\mathrm{acc}=\frac{GM}{c^2}, 
\end{equation}
$c$ being the sound velocity. $R_\mathrm{H}$ is the Hill radius (Eq.
(\ref{eq:hill})). The external boundary conditions for the temperature and 
density of the
envelope are those corresponding to the nebular gas, $T$ and 
$\rho$ (see Sect. \ref{sec:disc}). For further details on this aspect, 
we refer the reader to Benvenuto \& Brunini (\cite{bb}).

Regarding the equation of state (EOS), we employ results presented in Saumon et
 al. (\cite{saumon}). For the grain opacities, we use the tables from Pollack 
et al. (\cite{pollackop}). For temperatures above $10^3 \; \mathrm{K}$ we 
consider Alexander \& Ferguson (\cite{alexander}) molecular opacities which are
available up to $T \leq 10^4 \; \mathrm{K}$ and for higher temperatures we 
consider opacities by Rogers \& Iglesias (\cite{rogers}).

\section{Results}
\label{sec:results}

\subsection{Comparison with previous studies}
\label{sec:comparison}

\begin{figure}
\centering
\includegraphics[angle=-90, width= 0.45\textwidth]{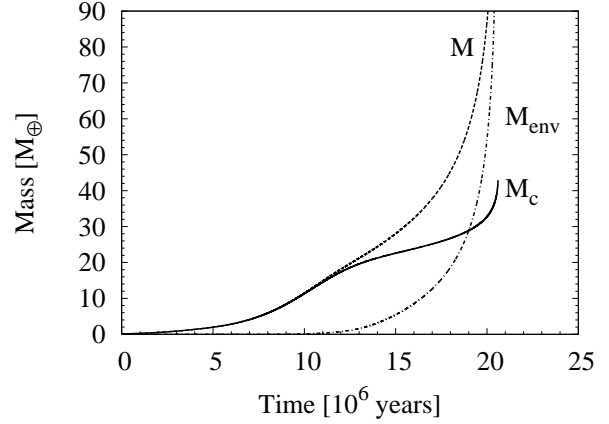}
\caption{Comparison case. We plot the mass evolution of the core
(solid line), of the envelope (dash--dotted line) and of the total mass (dashed 
line) of the protoplanet. The input parameters of this run are the same as those
of case J3 of Pollack et al. (\cite{pollack}). This figure should be compared to
Fig. 2b of the cited article. See the text for details.}
\label{j3m}
\end{figure} 
To our knowledge, no self--consistent calculation of giant
planet formation with a core growing according to the oligarchic model has been
 performed to date.
To show the impact of the core accretion rate on the whole formation of a
planet we performed a simulation to be compared with previous results
obtained by authors adopting another core accretion model. 
 
For comparison we selected one of the simulations performed by Pollack et al.
(\cite{pollack}), their case J3 for the formation of a Jupiter--like object.
These authors adopted for the growth of the core the model of Lissauer
(\cite{lissauer87}), which assumes a more rapid growth regime than the model of Ida
\& Makino (\cite{ida}) adopted here.
For this run, the protoplanetary disc at 5.2 AU and the accreted planetesimals 
are characterised as follows:
\begin{itemize}
\item initial solids surface density $\Sigma = 15 \; \mathrm{g \;cm^{-2}}$,
\item nebula volume density $\rho = 5 \times 10^{-11} \;
\mathrm{g \; cm^{-3}}$ ,
\item nebula temperature $T = 150 \; \mathrm{K}$,
\item planetesimal bulk density  $\rho_m=1.39 \; \mathrm{g \; cm^{-3}}$,
\item planetesimal radius  $r_m=100 \; \mathrm{km}$. 
\end{itemize}
The initial mass of the core, $M_\mathrm{c}$, is $0.1 \; \mathrm{M_{\oplus}}$ 
and the core density, $\rho_\mathrm{c}$, is $3.2 \; \mathrm{g \; cm^{-3}}$. 

\begin{figure}
\centering
\includegraphics[angle=-90, width= 0.45\textwidth]{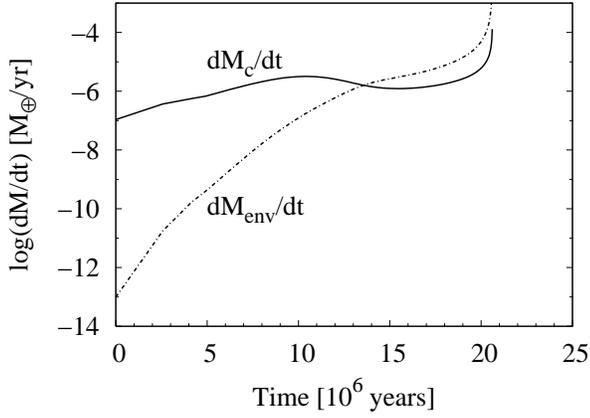}
\caption{For the same comparison case as in Fig. \ref{j3m}, solid 
(dashed--dotted) line shows the time evolution of the logarithm  of the core 
(envelope) growth rate. }
\label{j3mdot}
\end{figure}

The resulting core and envelope mass evolution of our run is depicted in Fig. 
\ref{j3m}, which should
be compared with Fig. 2b of Pollack et al. (\cite{pollack}). We use the same
mass range in the Y axis to favour the comparison, although the run was
completed successfully to the end, that is, until the total mass of the
planet, $M$, is that of Jupiter ($318 \; \mathrm{M_{\oplus}}$). From Fig.
\ref{j3m} it can be seen that in our calculation it 
takes 19 My for the mass of the envelope to equal the mass of the core, 
while in Pollack et al. (\cite{pollack})
the cross--over point ($M_\mathrm{c}=M_\mathrm{env}$) is reached in only 1.51 My. 
As is immediately obvious, the time difference between both calculations is 
of approximately one order of magnitude. This is a significant difference if 
we keep in mind that the estimated lifetime of the solar nebula is less than 10 
My. However, it is worth pointing out that the
cross--over mass in both simulations is almost the same ($\sim 29$ $\mathrm
{M_{\oplus}}$). For the sake of completeness, we mention that the whole formation 
time is 20.6 My and the final mass of the core is $42$ $\mathrm{M_{\oplus}}$. 
From the comparison of both figures, it can also be seen that we do not find
three phases in the formation of the planet as Pollack et al. (\cite{pollack})
found in their simulations. The slow growth of the core guarantees a smooth
variation of the slope of the $M_\mathrm{c}$ curve. No different 
growth regimes in the mass of the core or in the mass of the envelope can be 
distinguished before
the runaway growth of the envelope. Moreover, in Fig. \ref{j3mdot} we show in a
logarithmic scale, the solids and the gas accretion rate. Pollack et al. 
(\cite{pollack}) obtained a different behaviour of the accretion rates which
allowed them to define a short phase 1 in the growth of the protoplanet which
ends when $dM_\mathrm{c}/dt = dM_\mathrm{env}/dt$, and a longer phase 2 which 
ends when
$M_\mathrm{c}=M_\mathrm{env}$. Phase 2 is characterised by a constant
proportionality between $dM_\mathrm{c}/dt$ and $dM_\mathrm{env}/dt$. Phase 3
corresponds to the usual runaway growth of the envelope. From Fig. \ref{j3mdot} 
we can see that a distinction between phase 1 and a phase 2 cannot  
be inferred from our simulation. The time derivative of the mass of the 
envelope is a monotonously increasing function, with no flat slope after it 
becomes larger than the core growth rate.

\begin{figure}
\centering
\includegraphics[angle=-90,width= 0.45\textwidth]{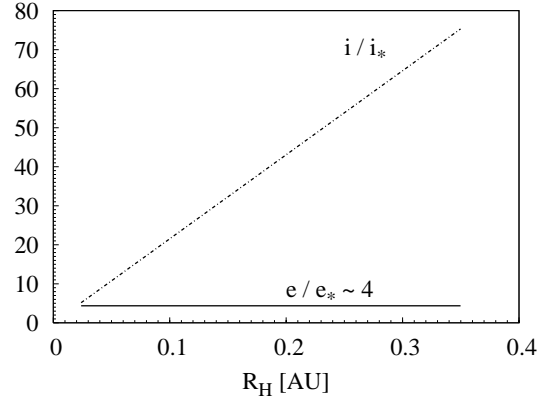}
\caption{The impact of the selected model for the growth of the core. The solid
line shows the ratio of eccentricities $e/e_*$, where $e$ is calculated 
according to Eq. (\ref{eq:excent}) and $e_*$ according to the model adopted by
Pollack et al. (\cite{pollack}). The dashed--dotted line plots the ratio of 
inclinations $i/i_*$. While in our model $i$ depends on $e$ ($i=e/2$), 
in Pollack et al.(\cite{pollack}) $i_*$ remains constant during the entire 
simulation. For the X axis we selected the Hill radius in order to get rid of 
the different formation timescales of both simulations.}
\label{j3ei}
\end{figure} 

Our physical model is in essence the same as that of Pollack et al.
 (\cite{pollack}), except for the selection of the core accretion model and the 
treatment of the interaction between incoming planetesimals and the
gaseous envelope of the protoplanet. Regarding  the latter, as explained in
Sect. \ref{sec:interaction}, we consider a simple model for the
energy deposition, while Pollack et al. (\cite{pollack}) developed a more 
complex model where the complete trajectories to the core of accreted 
planetesimals are calculated and
planetesimal ablation is also taken into account. However, this can not be the
main cause of such an important difference between the two formation timescales.
Although this effect may be important in determining the internal structure of
the protoplanet and may also modify the formation timescale, for this to happen
a significant amount of gas bound to the core is needed. In the present
simulation, it takes 10 My to form a core of $10 \; \mathrm{M_{\oplus}}$, 
and the mass of the envelope at that stage is only $0.1 \; \mathrm{M_{\oplus}}$. 
Hence, the main cause of the significant discrepancies between the timescales, 
must be the growth model for the core.
 We adopt an oligarchic growth for the core (with an enhanced 
effective radius due to the gas drag of the envelope), while Pollack et al.
(\cite{pollack}) prescribed the more rapid 
accretion model of Lissauer (\cite{lissauer87}), also modified by the
 enhancement caused by the gas drag. According to their Eq. (1), the core
 accretion rate can be written as:
\begin{equation}
\label{eq:p96acc}
\frac{dM}{dt} = \pi R^2 \Sigma \Omega F_\mathrm{g}
\end{equation}
where $R$ is the effective radius, $\Omega$ is the Keplerian angular velocity 
and $F_\mathrm{g}$ is the gravitational enhancement factor. 
 $F_\mathrm{g}$ depends on 
the RMS values of the reduced eccentricities and inclinations, and on the 
reduced effective radius,
\begin{displaymath}
e_{\mathrm{h},m}\equiv \frac{a}{R_\mathrm{H}} e \qquad i_{\mathrm{h},m} \equiv
\frac{a}{R_\mathrm{H}} i \qquad d_c \equiv \frac{R}{R_\mathrm{H}}.
\end{displaymath}
They calculate the numerical values of $F_\mathrm{g}$
 using the formulae obtained by Greenzweig \& Lissauer
(\cite{greenzweig}). Their model assumes that planetesimals' inclinations 
depend only on
planetesimal-planetesimal interactions and adopt for $i_{\mathrm{h},m}$ the 
formula:
\begin{equation}
\label{eq:ihp96}
i_{\mathrm{h},m}= \frac{v_{\mathrm{e},m}}{\sqrt{3} \Omega R_\mathrm{H}}
\end{equation}
where $v_{\mathrm{e},m}$ is the escape velocity from the surface of a 
planetesimal. Clearly, the inclination $i_*= i_{\mathrm{h},m} \; R_\mathrm{H}/a$ 
remains constant during the
entire calculation while $i_{\mathrm{h},m}$ decreases with time. On the other 
hand, they
assume that eccentricities are controlled by both planetesimals and protoplanet
 stirring, and prescribe for the reduced eccentricity the equation
\begin{equation}
\label{eq:ehp96} 
e_{\mathrm{h},m}= \mathrm{max} (2 i_{\mathrm{h},m},2). 
\end{equation}
This means that if 
$e_{\mathrm{h},m}=2 i_{\mathrm{h},m}$ the
protoplanet is growing according to the runaway regime, as $e$ and $i$ would be 
independent of
the mass of the protoplanet, while if $e_{\mathrm{h,m}}=2$ the eccentricity of
planetesimals would be affected by the presence of the protoplanet, this
condition corresponding to the 
protoplanet--planetesimal scattering in the shear--dominated regime. The
numerical simulations of Ida \& Makino (\cite{ida}) showed that the
protoplanet-planetesimal scattering in the shear--dominated regime lasts 
for only a few thousand years, after which  planetesimals are strongly 
perturbed by the protoplanet, and the post--runaway regime can then be 
considered as a dispersion--dominated regime. In the dispersion--dominated 
regime eccentricities and inclinations satisfy that  
$e_{\mathrm{h},m} / i_{\mathrm{h},m} \simeq 2$ 
and $e_{\mathrm{h},m} > 2$.
Hence, in a model where the protoplanet--planetesimal scattering 
is in the shear--dominated regime,  eccentricities and
inclinations of planetesimals in the vicinity of the protoplanet remain low.
This leads to an accretion scenario which is much faster than that corresponding
 to the oligarchic regime.

To show the difference between eccentricities and inclinations in both models
throughout the formation of the planet, we included in our simulation of the 
case J3, 
the calculation of $i_{\mathrm{h},m}$ and $e_{\mathrm{h},m}$ according to Eqs.
(\ref{eq:ihp96}) and (\ref{eq:ehp96}). The ratio of the eccentricity and
inclination between the two models is shown in Fig. \ref{j3ei}. We chose the 
X axis to be the Hill radius of the protoplanet, as it is
completely independent of time and the comparison is then straightforward. 
For this case, the
initial value of $i_{\mathrm{h},m}$ is lower than 1 and, as it is a decreasing 
function of the mass of the protoplanet, $e_{\mathrm{h},m}$ always equals 2. 
Consequently, our eccentricities are larger than those 
in Pollack et al. (\cite{pollack})
 by a factor of 4, as can be seen from Fig. \ref{j3ei}. 
On the other hand, our inclinations are
much higher (their inclination is constant throughout the whole formation of the 
planet, $i_* \simeq 0.0039$), which impacts not only in the relative velocities
 but also in
the scale height of the planetesimal disc which is inversely proportional to the
core accretion rate. As a consequence, due to our larger eccentricities and
inclinations, we obtain a much lower accretion rate and a 
longer formation timescale.

Pollack et al. (\cite{pollack}) adopt for the definition of the boundaries of 
the feeding zone that the radial distance on either side of the orbit is given 
by $\Delta a= \sqrt{12+e_{\mathrm{h},m}^2} R_\mathrm{H}$. According to our
simulation, $e_{\mathrm{h},m}$ is always equal to 2, which leads to 
$\Delta a= 4 R_\mathrm{H}$, the same value we adopted in our calculations.

\subsection{Results of full calculations of giant planet formation with an 
oligarchic growth for the core}
\label{sec:our_results}
All the simulations presented in this section were started with an 
embryo of $0.005 \; \mathrm{M_{\oplus}}$, revolving in a circular orbit 
around the Sun. The
semi--major axis of the orbit is that corresponding to Jupiter, $5.2$ AU. The
mass of the seed was chosen to guarantee that the embryo is undergoing an
oligarchic growth and that it is also able to bind a tiny atmosphere, the latter
in order to allow the code to converge. The density of the core will be held 
constant during the entire formation process, $\rho_\mathrm{core}=3 \; 
\mathrm{g \; cm^{-3}}$. The discrepancy between core density and planetesimals 
bulk density ($\rho_m=1.5 \; \mathrm{g \; cm^{-3}}$) is due to the progressive 
high pressure the core is subject to.
We arbitrarily stopped calculations when the total mass of the protoplanet was
that of Jupiter. This assumption tacitly implies that the nebula can always 
provide the necessary amount of gas for the formation of the planet.

The first set of simulations is intended to show the consequences of the
envelope's drag on the oligarchic growth of the core (Eq. (\ref{eq:accrete})).
To this end, we performed several runs for various protoplanetary disc surface
densities, including and excluding the effect of the gas drag 
in the calculation of $R_\mathrm{eff}$. The 
planetesimal radius is assumed to be constant, fixed for these examples at 100 
km (no size distribution or collisions between planetesimals were 
taken into account). As no planetesimal migration, planetesimal ejection by the  
protoplanet, perturbations by other embryos, etc., were considered,
 the variation of the solid surface density of the feeding zone is only due to 
planet accretion. Since no mechanisms for the dissipation of the
 gaseous component of the nebula were modelled, the variation of the 
 volume gas density of the feeding zone is due only to the formation of the 
envelope. We note that this is an idealised situation where the 
protoplanet probably has the largest amount of available material to feed 
itself. 
\begin{figure}
\centering
\includegraphics[angle=-90, width= 0.45\textwidth]{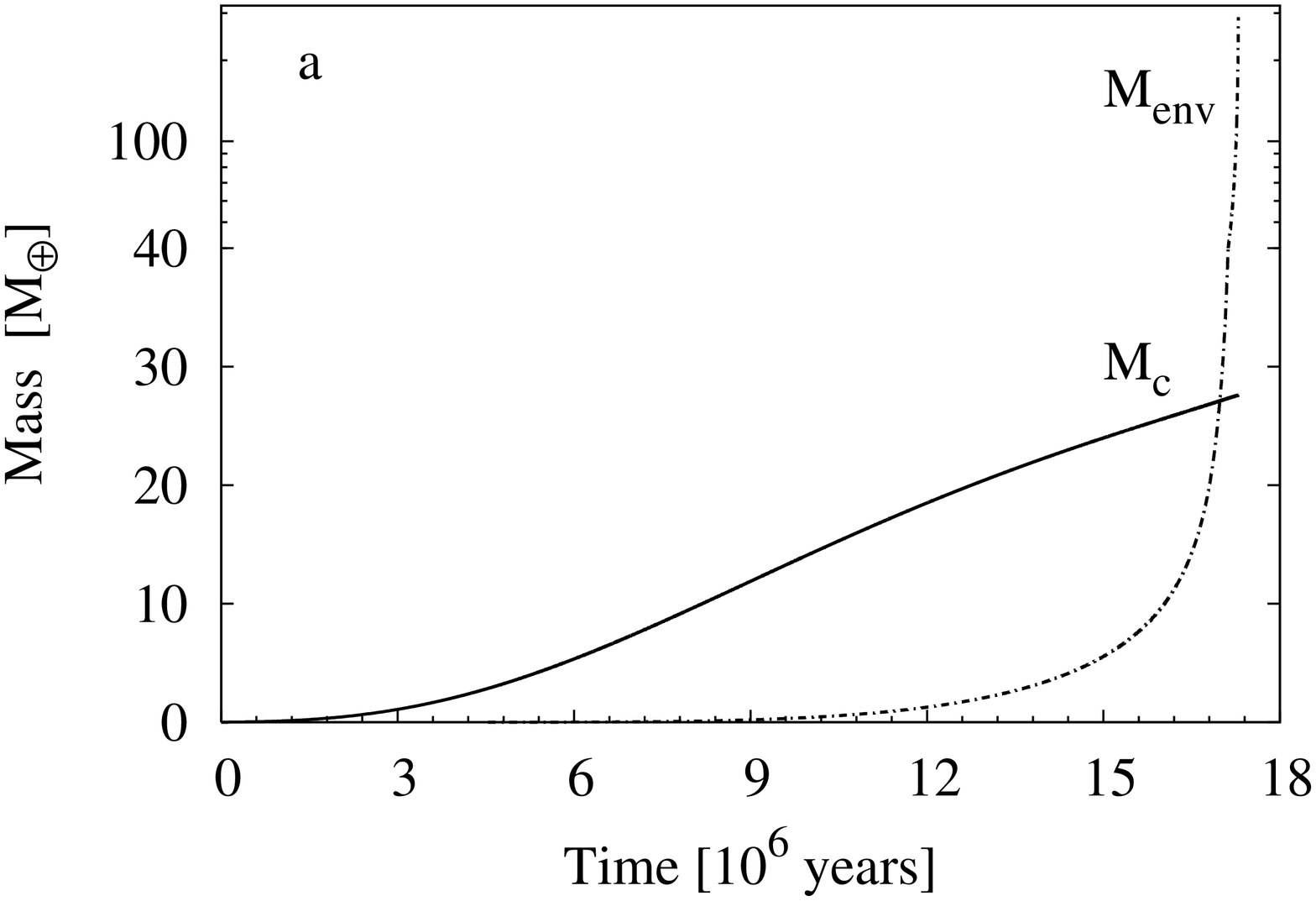}
\includegraphics[angle=-90, width= 0.45\textwidth]{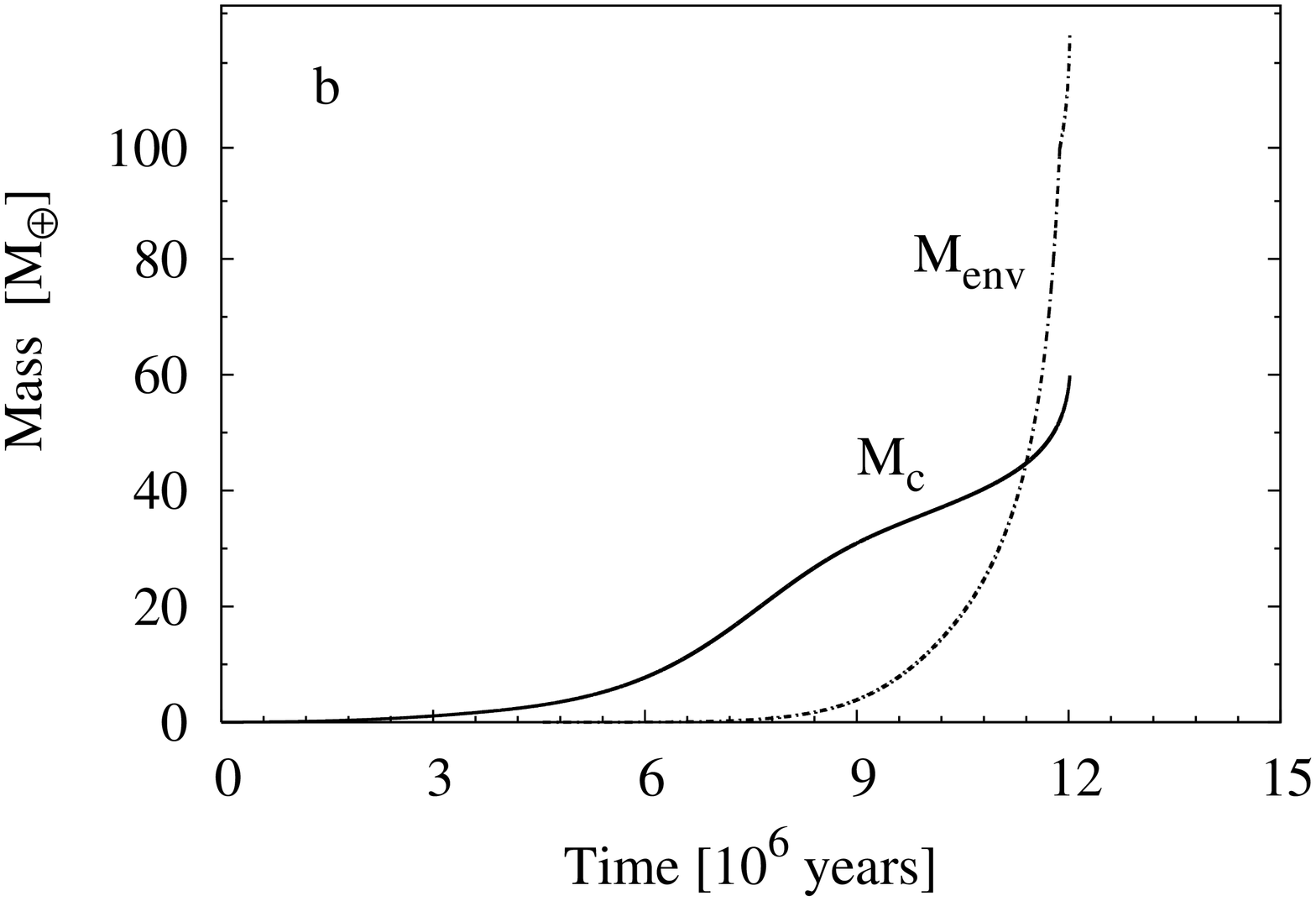} 
\caption{For a 6 MMSN protoplanetary disc, panel {\bf a} shows the core 
(solid line) and envelope (dashed--dotted line) mass evolution of a Jupiter--like
forming protoplanet at 5.2 AU, considering that the core grows according to the
oligarchic growth regime, with $R^*_\mathrm{eff}$ calculated with Eq.
(\ref{eq:reff}). Panel {\bf b} shows the same simulation but including the gas 
drag effect of the protoplanetary atmosphere in the calculation of 
$R_\mathrm{eff}$. Note that for a clear visualisation the Y axis changes from
a linear scale to a logarithmic scale. Panel {\bf a} ({\bf b}) is in a linear
scale up to $40 \; \mathrm{M_{\oplus}}$ ($100 \; \mathrm{M_{\oplus}}$), 
afterwards
the scale is logarithmic. For both runs, planetesimal radius was set equal to 
100 km.}
\label{comp1}
\end{figure} 

 We found that, in spite of the favourable conditions for mass accretion of our
model, for protoplanetary discs with densities lower than that of a 6 MMSN, 
the formation process 
could not be completed according to the timescales imposed by the 
observations of circumstellar discs (lower than $10^7$ yr) in either of the
cases considered for the calculation of $R_\mathrm{eff}$. 
The results for a disc of 6 MMSN are depicted in 
Fig. \ref{comp1}. The upper panel (Fig. \ref{comp1} {\bf a}) shows the 
evolution of the core and envelope mass when ignoring for the protoplanet 
capture effective radius the presence of the atmosphere ($R^*_\mathrm{eff}$ 
calculated according to Eq. (\ref{eq:reff})). The complete formation of a 
Jupiter--mass 
object takes a bit over 17 My and the final mass of the core is $\simeq$ 28
$\mathrm{M_{\oplus}}$ (note that, in our model, all accreted solids are 
deposited onto the core). However, when including the atmospheric gas drag, the 
timescale turns out to be 12 My (still over the limiting 10 My) while the mass 
of the core increases to 60 $\mathrm{M_{\oplus}}$ (see Fig. \ref{comp1} {\bf b}). 
This means that, in this case, the effect of the gas drag of the atmosphere 
reduces the formation time in about 30\%, but also affects the final mass of the
 core, which is increased a factor of 2. 
 \begin{figure}
\centering
\includegraphics[angle=-90, width= 0.45\textwidth]{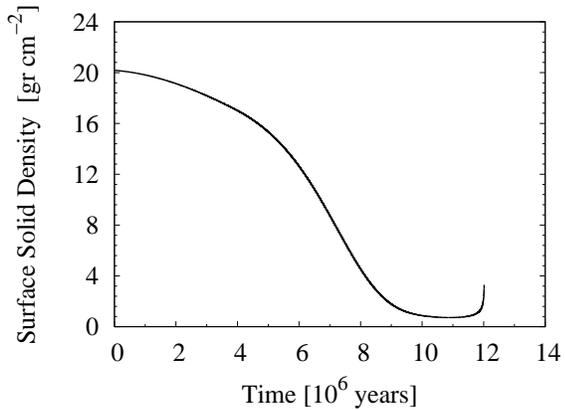}
\caption{This panel depicts the evolution of the solids surface density for the 
simulation that includes the atmospheric gas drag (corresponding to 
Fig. \ref{comp1} {\bf b}). A core of $\sim 10 \; \mathrm{M_{\oplus}}$ is 
achieved for $t \sim 6. 5 \; \mathrm{My}$, when $\Sigma$ is still very high.}
\label{density}
\end{figure} 

\begin{figure}
\centering
\includegraphics[angle=-90, width= 0.45\textwidth]{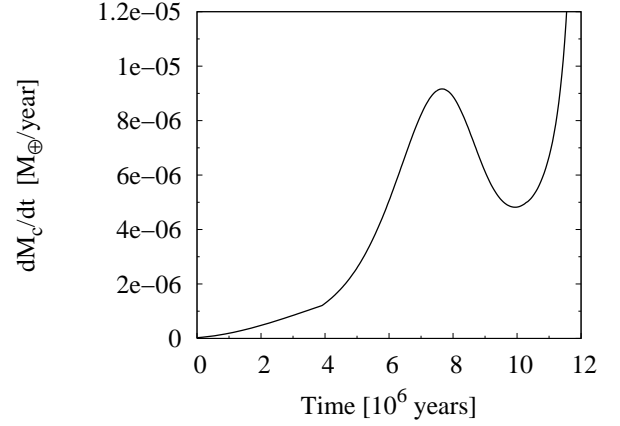}
\caption{This figure shows the evolution of the core accretion rate for the 
simulation that includes atmospheric gas drag. Note that the accretion rate 
is still an increasing function when the mass of the core reaches 10 
$\mathrm{M_{\oplus}}$ (corresponding to $\sim \mathrm{6.5 \; My}$;  see also 
Fig. \ref{comp1} {\bf b}), preventing the rapid accretion of the envelope mass.}
\label{mc}
\end{figure} 

Although a core of $\sim 10 \; \mathrm{M_{\oplus}}$ (currently an acceptable 
value for Jupiter's core mass, Saumon \& Guillot \cite{guillot}) is formed in 
$\sim \, \mathrm{6.5 \; My}$ in the second simulation, the runaway collapse of 
the gaseous envelope occurs $ \sim 5$ My later. As seen from Eq. 
(\ref{eq:accrete}), the solids accretion rate is directly proportional to the 
solids surface density  and the fact that the feeding zone is far from being 
depleted when 
this mass is achieved (see Fig. \ref{density}), allows the protoplanet continue  
the accretion of solids. In fact, as seen from Fig. \ref{mc}, 
the 
peak of the core accretion rate takes place  for $t \, \sim 8$ My, when the mass
 of the core is approximately 20 $\mathrm{M_{\oplus}}$. 
This large number of incoming planetesimals contribute to the gas pressure
supporting the envelope via their gravitational energy, and thus prevent faster
accretion of gas from the nebula and as such permit the formation of a massive
core.

\begin{centering}
\begin{table}
\caption{ \label{table:times} Comparison of formation times and core 
masses for several disc densities. Planetesimal radius is 100 km.}
\begin{tabular}{l p{1.3 cm} p{1 cm} p{1.3 cm} p{1 cm}}
\hline
Disc density  & \multicolumn{2}{c}{Without gas drag} &  \multicolumn{2}{c}
{With gas drag}\\
\cline{2-5}
[MMSN] & Formation time [My] & Core mass [$\mathrm{M_{\oplus}}$] & 
Formation time [My] & Core mass [$\mathrm{M_{\oplus}}$]\\
\hline
\\
$\; \; 6$   & 17 & 28  & 12  & $\; \; 60$ \\
$\; \; 7$   & 14 & 32  & $\; \; 9$  & $ \; \; 70$ \\
$\; \; 8$   & 11 & 36  & $\; \; 7$  & $\; \;  81$ \\
$\; \; 9$   & 10 & 40  & $\; \; 6$ & $\; \;  91$ \\
10  & $\; \; 8$ & 43  & $\; \; 5$  & 100 \\
\hline 
\end{tabular}
\end{table}
\end{centering}

\begin{centering}
\begin{table}
\caption{ \label{table:times2} Comparison of formation times and core 
masses for several disc densities with a planetesimal radius of 10 km.}
\begin{tabular}{l p{1.3 cm} p{1 cm} p{1.3 cm} p{1 cm}}
\hline
Disc density & \multicolumn{2}{c}{Without gas drag} &  \multicolumn{2}{c}{With gas
 drag}\\
\cline{2-5}
 [MMSN]& Formation time [My] & Core mass [$\mathrm{M_{\oplus}}$] & 
 Formation time [My] & Core mass [$\mathrm{M_{\oplus}}$]\\
\hline
\\
$\; \; 6$   & 9   & 34  & 5.6  & $\; \; 69$ \\
$\; \; 7$   & 7   & 40  & 4    & $\; \; 80$ \\
$\; \; 8$   & 6   & 45  & 3    & $\; \; 92$ \\
$\; \; 9$   & 4.8 & 50  & 2.2  &  104 \\
10  & 4   & 55  & 1.75 & 115 \\
\hline 
\end{tabular}
\end{table}
\end{centering}

The same behaviour is found when increasing the surface density of the disc. 
In Table 
\ref{table:times} we show, in round numbers, the formation times and the 
final masses of protoplanetary cores for discs that range from 6 MMSN 
to 10 MMSN and planetesimals having a radius of 100 km. As expected, the higher 
the density of the disc, the shorter the formation time but the larger the final 
core's mass (see also Pollack et al. \cite{pollack}). The timescale 
 reduction, when the gas drag from the atmosphere is included, ranges 
from $\sim 30\%$ for 6 MMSN to $\sim 40 \%$ for 10 MMSN. The maximum 
timescale reduction for these cases is then less than a factor of 2. The mass 
of the 
core, on the other hand, monotonously increases fractionally, from a factor of 
$\sim$ 2.15 for 6 MMSN to a factor of $\sim$ 2.35 for 10 MMSN. Qualitatively,
all curves look similar to that shown in Fig. \ref{comp1} and it is not worth 
showing them here.

When the same comparative simulations are made for planetesimals having a radius
of 10 km (this means, simulations that consider both cases for the 
effective radius of the protoplanet, $R^*_\mathrm{eff}$ and 
$R_\mathrm{eff}$), the timescale reduction ranges from $\sim 38 \%$ 
for 6 MMSN to $\sim 56 \%$ for 10 MMSN, and again we find that the mass 
of the core is doubled (see Table \ref{table:times2}).  For planetesimals of 
this size,
and according to the results presented in Table \ref{table:times2}, one could 
speculate about reducing the density of the disc (to less than that
corresponding to 6 MMSN) and still finding 
 a formation timescale below $10^7$ yr (when considering the 
enhanced capture radius by gas drag).
However, to present our results clearly, we find it more 
relevant to compare the several sets of simulations for the same nebula
conditions.

\begin{figure}
\centering
\includegraphics[angle=-90, width= 0.45\textwidth]{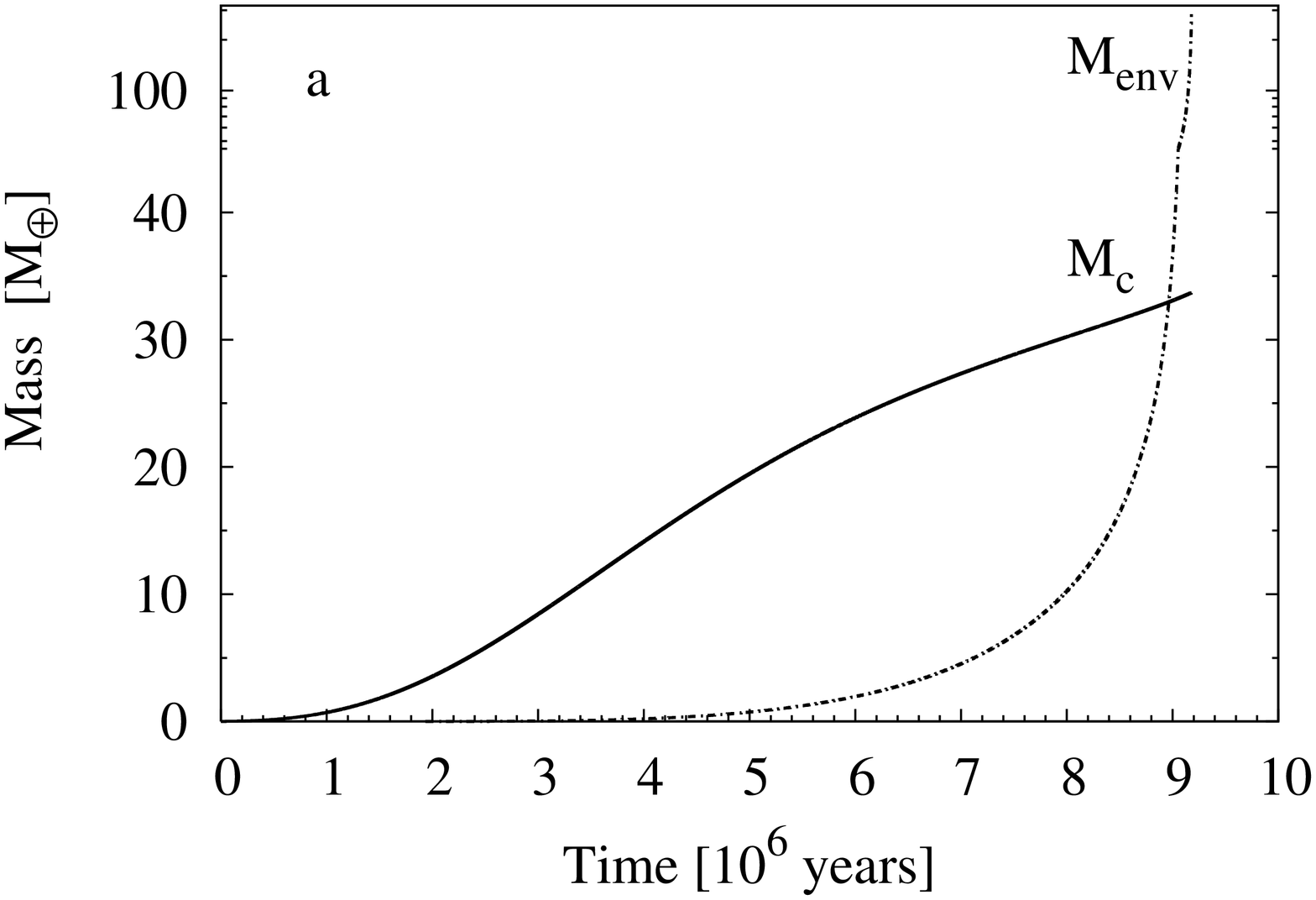}
\includegraphics[angle=-90, width= 0.45\textwidth]{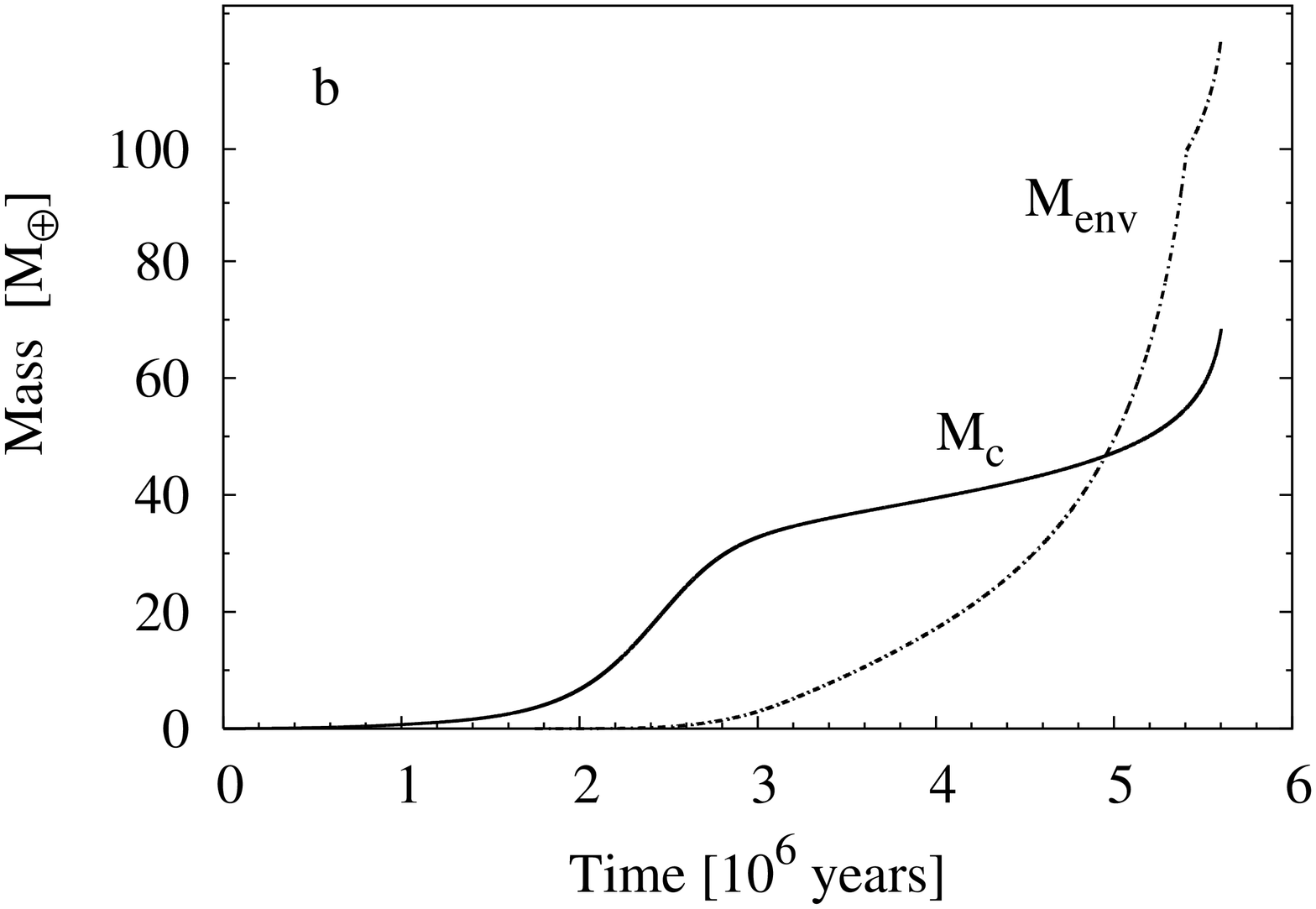} 
\caption{For a protoplanetary disc of 6 MMSN and planetesimals having a radius 
of 10 km, upper panel ({\bf a}) shows the evolution of the protoplanet mass 
(the core mass with the solid line, the envelope mass with the dashed--dotted 
line) when no atmospheric gas drag is considered. 
The lower panel ({\bf b}) illustrates the same 
process but when $R_\mathrm{eff}$ is calculated including the gas drag 
force acting on incoming planetesimals. Again, the Y axis is split in two
different scales: first we adopt a linear scale and then a logarithmic one. The
same range as in Fig.~(\ref{comp1}) is adopted.}
\label{comp2}
\end{figure}

For the sake of completeness, Fig. \ref{comp2} shows the evolution of the core 
mass and the envelope mass for a 6 MMSN disc where the planetesimal radius is 
10 km.
When gas drag from the envelope of the protoplanet is ignored, the 
formation time is $\sim$ 9 My and the core mass is $\sim 34 \;
\mathrm{M_{\oplus}}$. However, when this effect is included in the calculations, 
the total formation time is reduced to $\sim 6 \; \mathrm{My}$, with a 
final super--massive core of $69 \; \mathrm{M_{\oplus}}$.

Although our core masses are in all cases completely out of range for present 
models of Jupiter total mass of solids, $M_{\mathrm{Z}}$ (recent estimations are
$M_{\mathrm{c}} \la 11 \; \mathrm{M_{\oplus}}$ and $1 \; \mathrm{M_{\oplus}}
 \la M_{\mathrm{c}} + M_{\mathrm{Z}} \la 39 \; \mathrm{M_{\oplus}}$, 
Saumon \& Guillot \cite{guillot}), the main objective of this work is not 
directed to fit our results to these values, but on stressing the results
obtained when adopting the oligarchic
growth for the core. From our calculations it is clear that including the
enhancement of the accretion cross--section due to gas drag, helps to reduce the
formation timescale when considering an oligarchic model, 
although surface densities higher than that of the MMSN
are still needed. The main drawback of these massive discs is that they lead to
the formation of super--massive cores. However, we have to keep in mind that no
mechanisms of planetesimal removal from the feeding zone of the protoplanet are
taken into account, except for the depletion due to planetesimal accretion onto
the protoplanet.

Thommes et al. (\cite{thommes}) calculated analytically and numerically the
evolution of solid protoplanetary masses growing oligarchical for a wide 
range of semi--major axes and for a protoplanetary disc of 1
MMSN and 10 MMSN (their solids surface density for the MMSN is slightly lower 
than ours). For a 10 MMSN and considering planetesimals having a radius of 10 km, 
they obtain a protoplanet of $\simeq 60 \,\mathrm{M_{\oplus}}$, a
result we were also able to reproduce in the absence of gas accretion. However, 
when they include in their simulations the planetesimal migration in the
protoplanetary disc due to nebular gas drag, the resulting protoplanet is 
reduced to 
$\sim 10 \;\mathrm{M_{\oplus}}$. Planetesimal migration due to gas drag is a 
relevant ingredient which should be considered in future calculations for an 
accurate estimation of the mass of the core and of the formation timescale.
Another important effect that should be taken into account in future 
calculations is the ejection of planetesimals out of the feeding zone. This
could probably also help in solving the problem of having a Jupiter--like planet
with such large core masses. Since ejection occurs mainly for large mass
planets, this would not necessarily slow down the accretion of planetesimals at
the beginning, and may not lead to a too long formation time of the planet.

The previous paragraphs focus on the consequences of the 
enhancement of the effective capture cross--section due to the envelope's gas 
drag. We now analyse the importance of planetesimal size. 
For a clear visualisation we summarise in Table \ref{table:comp2}
the earlier calculations with the enhanced 
 effective radius for planetesimals having radii of 10 and 100 km.
 Planetesimal mass appears in Eq. 
(\ref{eq:excent}), which determines the relative velocity of planetesimals with
respect to the protoplanet through Eq. (\ref{eq:vrel}). Smaller planetesimals 
experience stronger damping of random velocities due to nebular gas and form a 
thinner disc ($i \simeq \frac{1}{2} \, e$ and $h \simeq  a \,i$) which in turn  
favours accretion (Thommes et al. \cite{thommes}). 
Moreover, smaller planetesimals 
suffer from a stronger deceleration due to atmospheric gas drag, increasing 
the impact parameter. Both effects combined favour the formation 
process, reducing the formation timescale.

If we analyse and compare results without including the gas drag of the
envelope for a fixed nebula density  
(first and second Cols. of Tables \ref{table:times} and 
\ref{table:times2}), 
we find that for planetesimal radius of 10 km the whole formation process 
is reduced in an approximately constant factor of 2 when compared to
planetesimals having a radius of 100 km, while the core mass is increased by
 \mbox{25\%}. This value is independent of the protoplanetary disc
 conditions,
 at least in the cases considered here. 
When the gas drag is included in the calculations this behaviour changes (see
Table \ref{table:comp2}). 
Results show that the formation time reduces from a factor of $\sim 2$ for a 
protoplanetary disc of 6 MMSN to a factor of $\sim 3$ in the case of 10 MMSN,
when comparing the runs for the same protoplanetary disc conditions but for 
different planetesimal size.
The factor of the timescale reduction depends then, on the surface density  of 
the protoplanetary disc.
The mass of the core is less affected than in the previous case, resulting for
planetesimals having a 
radius of 10 km, as being on average, \mbox{15\%} more massive than for planetesimal 
radius of 100 km. These results show that when including the gas 
drag of the envelope, the reduction factor in
the formation time also depends on the surface density of the disc.
 Note that the fractional decrease of the whole 
formation time is greater than the fractional increase of the core mass.

\begin{centering}
\begin{table}
\caption{\label{table:comp2} Comparison of formation times and core masses
 for several disc densities, and for two different planetesimal sizes: 100 and 
 10 km. All cases were calculated including atmospheric gas drag.}
\begin{tabular}{p{1.0cm} p{1.2cm} p{1cm} p{1.2cm} p{1cm}}
\hline
Disc density & \multicolumn{2}{p{1.9cm}}{Planetesimal radius: 100 km} &  
\multicolumn{2}{p{1.8cm}}{Planetesimal radius: 10 km}\\
\cline{2-5}
 [MMSN] & Formation time [My] & Core mass [$\mathrm{M_{\oplus}}$] & Formation 
 time [My] & Core mass [$\mathrm{M_{\oplus}}$]\\
\hline
\\
$\; \; 6$   & 12  &  $\; \; 60$  & 5.6  & $\; \;69$ \\
$\; \; 7$   & $\; \; 9$  & $\; \; 70$  & 4    & $\; \; 80$ \\
$\; \; 8$   & $\; \; 7$  & $\; \; 81$  & 3    & $\; \; 92$ \\
$\; \; 9$   & $\; \; 6$  & $\; \; 91$  & 2.2  & 104 \\
10  & $\; \; 5$  & 100  & 1.75 & 115 \\
\hline 
\end{tabular}
\end{table}
\end{centering}

In summary, this section focused on the study of the formation of a giant 
planet with a full evolutionary code, 
assuming an oligarchic growth regime for the core. 
We first studied the enhancement of the effective capture radius due to the
presence of the gaseous envelope. The atmospheric gas drag enlarges the capture
cross--section, causing a reduction in the formation time but also a significant 
increment in the mass of the core. 
We also studied the impact of the size of accreted planetesimals in
our model.
We found that the whole formation process is accelerated when considering 
smaller planetesimals and that the reduction factor depends on the surface 
density of the feeding zone. 
However, no substantial change in the final mass of the core was identified.

\section{Discussion and summary }
\label{sec:conclusion} 

In this paper we examined the formation of a Jupiter--like object.
Motivated by the work of Thommes et al. (\cite{thommes}), we selected  
 an oligarchic growth regime for the core as the planetesimal accretion rate. 
 The numerical simulations were made
with the code presented in a previous paper by Benvenuto \& Brunini
(\cite{bb}), which was updated in this study (see Sect. \ref{sec:procedure}).
 This code was developed for a self--consistent 
calculation of giant planet formation, so the
growth of the core is coupled to the growth of the envelope. In this sense, 
the evolution of the 
envelope's density profile is a natural outcome of the code and a relevant 
quantity for an accurate estimation of the enhancement in the protoplanet 
capture cross--section of planetesimals (as the drag force depends on 
$\rho_\mathrm{g}$, see Eq. (\ref{eq:stokes})), which has a direct impact  
on the growth rate of the core.

When including a time--dependent planetesimal accretion rate in these kinds of 
calculations, most authors prescribe a rapid one, that in general 
guarantees the formation of a massive core in a few hundred thousand years. 
However, N--body simulations show that when the embryo is about the mass of 
the Moon, or even smaller (Thommes et al. \cite{thommes}), the growth regime
enters the protoplanet--dominated stage. For this 
reason, we assume from the very beginning of our calculations an oligarchic 
growth regime, which seems to be a more realistic approach to the problem. This 
accretion rate is considerably slower than the usually--adopted 
accretion model of Greenzweig \& Lissauer (\cite{greenzweig}).
 Adopting the oligarchic growth model has, as a main consequence,  an increase
 in the whole formation timescale  when compared to previous calculations
which adopt a rapid growth for the core (see Sect. \ref{sec:comparison}). 
In this sense, it is worth mentioning that an accurate giant planet formation
model with an oligarchic growth for the core should also include 
 planetesimal migration due to the nebular gas drag since 
the assumed lifetime of the nebula is of the same order of magnitude of the 
formation process itself (Thommes et al. \cite{thommes}). 
In the calculations presented here this effect was not included, as 
this article aims to analyse in a first approximation the oligarchic growth
regime for a giant planet in a self--consistent calculation.

To the authors' knowledge, the oligarchic growth of protoplanets has
generally been studied in the absence of an atmosphere and no 
full evolutionary calculations of gaseous giant planets adopting this 
growth regime for the core have been performed in the past. 
As giant planets have large gaseous envelopes, the gas drag effect of the 
growing atmosphere should play an important role in the resulting accretion 
cross--section of the protoplanet. In order to study the relevance of this 
effect, we made a set of simulations for several disc densities to estimate the 
consequences of including
the gas drag in the calculation of the effective radius and we compared them to 
those where the effective radius is calculated in the absence of gas drag. 
According to our simulations, the main effects of the enlargement of the
effective radius when considering the gas drag are a reduction of about 
\mbox{30\%} to \mbox{55\%} in the whole formation timescale 
 (depending on the surface density of the disc and on
the planetesimal mass, being stronger for smaller 
planetesimals) and an increase in the mass of 
the core by about a factor of 2 for the several protoplanetary disc 
considered here. As we mentioned before, no planetesimal migration or other 
protoplanet--disc interactions were considered, so our results should be 
analysed in the context of this simple model. However, from the comparison of 
simulations with and without atmospheric gas drag, we conclude that including 
the gas drag in the 
calculation of the effective cross--section of the protoplanet is absolutely 
relevant to the estimation of the formation timescale. The fact that, in 
association with higher surface densities for the disc, we obtain the desired 
shorter timescales but, on the other hand, super--massive cores should not be 
a cause for concern. Although the mass of the core is closely related to the 
formation timescale (the more massive the core, the shorter the timescale),
Thommes et al. (\cite{thommes}) showed in their calculations of the formation of
protoplanetary cores, that planetesimal migration strongly affects the 
planetesimal population of the feeding zone, being very effective in depleting 
it. Including this effect reduces considerably the final mass of the core,
and protoplanetary discs as massive as 10 MMSN would probably offer us 
good fits to the core mass and the formation timescale simultaneously.

In this paper, we also explored the effect on giant planet formation of 
planetesimal size variations. For 
the same set of protoplanetary discs (from 6 to 10 MMSN), we made 
simulations for a swarm of planetesimals having a fixed radius of 
10 and 100 km. We find 
that the formation timescale is strongly dependent on planetesimal size: for 
planetesimal radius of 10 km, the process occurs a factor of 2 to 3 times 
faster than for the case of larger planetesimals. A reduction in the 
timescale was expected, since smaller planetesimals have lower relative 
velocities which favour accretion and they are much more affected by gas 
drag. However, the mass of the core is not very increased.

In a recent paper, Inaba \& Ikoma (\cite{inaba}) also explored the effects
of gas drag on the effective radius of a protoplanet for capturing 
planetesimals of different sizes.
 They found that gas drag largely increases the 
effective capture radius. While our results are qualitatively similar to 
those of Inaba \& Ikoma (\cite{inaba}), we note that we employed 
a full evolutionary code. In particular, it is worth discussing the results 
Inaba \& Ikoma (\cite{inaba}) presented in their Fig.~6. They estimated 
the time spent in forming a $10 \; \mathrm{M_{\oplus}}$ core as a function 
of planetesimal size, finding that gas drag drastically decreases the 
formation timescale: for planetesimals of, e.g., 10 km (100 km) the 
timescale is reduced by a factor of $\sim 6$ ($\sim 4$). 
These factors would be very 
helpful in alleviating the timescale problem of the whole core growth  
mechanism. However, the results presented in this paper indicate that, 
while there exists a reduction in the time spent in forming a core of 
$10 \; \mathrm{M_{\oplus}}$, it is much more modest. We find that it is only 
slightly dependent on the protoplanetary surface density and 
moderately dependent on 
the planetesimal size. For planetesimals of 10 km the average reduction 
factor in the formation timescale of a $10 \; \mathrm{M_{\oplus}}$ core is 
$\sim 1.25$  and for planetesimals of 100 km it is 
$\sim 1.5$. It is also worth mentioning that the timescale involved in forming 
a $10 \; \mathrm{M_{\oplus}}$ core is not really representative of the complete
giant planet formation timescale as shown in this and  
previous papers (e.g., Pollack et al. \cite{pollack}).

The next step for our future calculations is the inclusion of planetesimal 
migration due to nebular gas drag and planetesimal ejection out of the feeding
zone. This will provide us with a more accurate quantitative idea of the final 
core
masses and formation timescales in the frame of an oligarchic growth regime for
the core. Also, a size distribution of the accreted planetesimals should 
be considered for a better estimation of the formation timescales. 
Or, at least, if a unique radius is adopted for all 
planetesimals in the swarm, several simulations varying the radius should be 
made in order to bracket the formation time. Finally, we expect to 
update the opacity tables since opacity plays a fundamental role 
in the formation timescale (Podolak \cite{podolak}; 
Hubickyj et al. \cite{hubickyj}). It would also be interesting to test with our 
model, the response of the whole planetary formation process to grain 
opacity reduction, which will probably shorten the timescales involved, as 
has been already shown by, e.g., Pollack et al. (\cite{pollack}) and 
Hubickyj et al. (\cite{hubickyj}).

\begin{acknowledgements}
 A.B. acknowledges the financial support by ANPCyT.  
   
 The authors wish to thank the anonymous referee for constructive 
criticisms which enabled us to largely improve the original version of this
 article.
\end{acknowledgements}

\end{document}